\documentclass[english]{newFNLstyle}

\usepackage[latin1]{inputenc}
\usepackage{url}
\usepackage{graphics,epsfig}

\makeatletter

\providecommand{\LyX}{L\kern-.1667em\lower.25em\hbox{Y}\kern-.125emX\@}

\usepackage{verbatim}

\usepackage{babel}
\makeatother

\begin{document}

\volnumpagesyear{0}{0}{000--000}{2002}
\dates{received date}{revised date}{accepted date}

\title{ANALYTICAL EXPRESSIONS FOR PARRONDO GAMES}

\authorsone{LARS RASMUSSON and MAGNUS BOMAN}
\affiliationone{Swedish Institute of Computer Science}
\mailingone{Box 1263, SE-164 29 Kista, Sweden \\
		lra@sics.se, mab@sics.se}

\maketitle

\markboth{Analytical Expressions for Parrondo Games}{Rasmusson and Boman}

\pagestyle{myheadings}


\keywords{Parrondo game, Parrondo's paradox, game theory, game state, optimal mixing strategy}

\begin{abstract}
We present some new analytical expressions for the so-called Parrondo
effect, where simple coin-flipping games with negative expected
value are combined into a winning game. Parrondo games are state-dependent.
By identifying the game state variable, we can compute the stationary
game state probabilities. Mixing losing games increases the probability
of entering into a game state with positive value, as can be seen
clearly from our analytical expressions for the Parrondo game value.


\end{abstract}

\section{\label{sect:intro}Introduction}

A \textit{Parrondo game} is a combination of two or more simple losing games in which possibly biased coins are flipped, a strategy for alternation between the simple games, and a state. The state determines the probabilities of winning in one or more of the simple games and codes the game history, including the capital of one or more players of the Parrondo game. A winning strategy yields a positive expected value of the Parrondo game, in spite of the constituent simple games having negative expected value: the so-called Parrondo effect (often referred to as Parrondo's paradox~\cite{haab02}). Conditions for the simple games and for the strategies (and implicitly also for the Parrondo game states) were given by Harmer and Abbott~\cite{haab99a}. We give new analytical expressions for results previously approximated by either computer simulations or discrete time Markov chain analyses~\cite{haab02,haabtapa01,kajo02,ra02}. 


The original Parrondo game rules~\cite{haab99} combined two games, of which the second was later modified to \textquotedblleft present new games where all the rules depend only on the history of the game and not on the capital\textquotedblright~\cite{pahaab00}. 
We begin by analyzing this modified game, named $B'$.
We perform the simple calculations reproducing the known result~\cite{pahaab00}
that the ergodic expected value of $B'$ is negative, in order to introduce the notation
used in subsequent sections.
In Section 3, it is shown that it is possible to obtain a winning game by adjusting the state transition probabilities in the game in which $B'$ is mixed with the original Parrondo game $A$. We then calculate the optimal strategy for the mixed game. Finally, we analyze the original Parrondo game in an analysis that requires the introduction of a new state parameter: the capital of the player.

\section{\label{sect:bprime}Game $B'$}

The set of possible outcomes of game $B'$ is $\Omega = \{-1,1\}$, also called
\textit{losing} and \textit{winning}. The game history $g_{t}\in \Omega$ is the
outcome of game $B'$ at time $t$. The probabilities of the outcomes
depend on the game history in the following way\begin{eqnarray}
p_{1|-1,-1} & = & 9/10-\epsilon \nonumber \\
p_{1|1,-1}=p_{1|-1,1} & = & 1/4-\epsilon \label{eq:args}\\
p_{1|1,1} & = & 7/10-\epsilon \nonumber 
\end{eqnarray}
where we use the notation \begin{eqnarray*}
p_{ijk\cdots }^{t} & \doteq  & \textrm{Prob}[g_{t}=i,g_{t-1}=j,g_{t-2}=k,\ldots ]
\end{eqnarray*}
for the time-dependent distribution, $p_{ijk\cdots }$ for the ergodic
distribution where \begin{eqnarray*}
p_{ijk\cdots }^{t}=p_{ijk\cdots }^{t-1}\textrm{, and}
\end{eqnarray*}
\[
p_{i|j\cdots }=\frac{p_{ij\cdots }}{p_{j\cdots }}\]
For example, the probability of winning after having lost two simple games
is $\textrm{9/10-}\epsilon $.


For the ergodic process, it holds that\begin{equation}
p_{ij}=\sum _{k\in \Omega}p_{i|jk}p_{jk},\qquad \sum _{(i,j)\in \Omega^{2}}p_{ij}=1\label{eq:2}\end{equation}
The linear system (\ref{eq:args}) and (\ref{eq:2}), has the following solution (cf.~\cite{haab02}), with the rightmost column indicating the value for $\epsilon$ = 0.001:
\[
\begin{array}{lcll}
 p_{-1,-1}^{*} & = & \left(45+210\epsilon +200\epsilon ^{2}\right)/C & 0.228\\
 p_{-1,1}^{*}=p_{1,-1}^{*} & = & \left(2(27+60\epsilon -100\epsilon ^{2})\right)/C & 0.273\\
 p_{1,1}^{*} & = & \left(45-230\epsilon +200\epsilon ^{2}\right)/C & 0.226\end{array}
\]
where $C=198+220\epsilon$. The ergodic expected value of the game is:

\begin{eqnarray}
\left\langle g_{t}\right\rangle  & = & \sum _{i\in \Omega}i\, p_{i}\nonumber \\
 & = & \sum _{(i,j,k)\in \Omega^{3}}i\, p_{i|jk}p_{jk}\nonumber \\
 & = & p_{1,1}(\frac{2}{5}-2\epsilon )+(p_{-1,1}+p_{1,-1})(-\frac{1}{2}-2\epsilon )+p_{-1,-1}(\frac{4}{5}-2\epsilon )\label{eq:3}\\
 & = & -\frac{20\epsilon }{9+10\epsilon }\label{eq:4}
\end{eqnarray}
Thus, the game has negative expected value for $\epsilon >0$.

\section{\label{sect:mix}Mixing Simple Games}

Mixed with another game, $B'$ can have a higher expected value because
the outcome probabilities $p^{*}$ of the mixed game depend on the Parrondo game history $g_{t}^{*}$
rather than the simple game history $g_{t}$. Eq. (\ref{eq:3}) shows
that if $\epsilon < 1/5$ and
$p_{1,1}^{*}\geq p_{1,1}$ and
$p_{-1,-1}^{*}\geq p_{-1,-1}$, then $\left\langle g_{t}^{*}\right\rangle \, \geq \, \left\langle g_{t}\right\rangle $, since $p_{1,1}$ and $p_{-1,-1}$ both have positive coefficients.

The original biased coin-flipping game $A$ has outcome $1$ with probability $q_{1}=1/2-\epsilon $,
and outcome $-1$ otherwise (see~\cite{haab99}), thus its expected value is $-2\epsilon$.  It was in~\cite{pahaab00} mixed with $B'$, leading to an interest in the \textit{mixing parameter}, here denoted by $u$. 
In the original game set-up, $u=1/2$ is the probability that $g_{t}^{*}$
is the outcome of game $A$,
otherwise it is the outcome of $B'$. The mixed game has positive
expected value, i.e. $\left\langle g_{t}^{*}\right\rangle >0$ for
some $\epsilon >0$. The fact the simple game $B'$ in this mixed game
also has positive expected value goes unremarked in~\cite{pahaab00}.
More specifically, the negatively biased coin-flipping original game increases
the probability $p_{-1,-1}^{*}$ for two consecutive losses
in the mixed game, which in turn increases the expected value of the game
$B'$ enough to compensate for the loss suffered from the other simple game.
For the mixed game $p_{ij}^{*}$ it holds that \begin{eqnarray}
p_{ij}^{*} & = & u^{2}q_{i}q_{j}\nonumber \\
&&+u(1-u)q_{i} (\sum _{k,l \in \Omega}p_{j|kl}p_{kl}^{*})\nonumber \\
&&+(1-u)u (\sum _{j,k \in \Omega}p_{i|jk}q_{j}p_{kl}^{*}) \label{eq:5} \\
&&+(1-u)^{2} (\sum _{j,k \in \Omega}p_{i|jk}p_{j|kl}p_{kl}^{*})\nonumber 
\end{eqnarray}
\begin{eqnarray}
\sum _{(i,j)\in \Omega^{2}}p_{ij}^{*} & = & 1\label{eq:6}
\end{eqnarray}
since $p_{ij}^{*}$ depends on both of the simple games.
The linear system Eq. (\ref{eq:args}), (\ref{eq:5}) and (\ref{eq:6})
has for $u=1/2$ the solution (with the rightmost column indicating the value for $\epsilon$ = 0.001)
\[
\begin{array}{lcll}
 p_{-1,-1}^{*} & = & 10(2+5\epsilon) (5+8\epsilon)/C & 0.234\\
 p_{-1,1}^{*}=p_{1,-1}^{*} & = & -8(2+5\epsilon) (-7+10\epsilon)/C & 0.261\\
 p_{1,1}^{*} & = & 5(3-8\epsilon) (7-10\epsilon)/C & 0.244\end{array}
\]
where $C=429+220\epsilon$.
This results in a positive expected value of the Parrondo game,
now parametrised with respect to $u$ as well as to $\epsilon$: \begin{eqnarray}
\left\langle g_{t}^{*}\right\rangle  & = & 
\frac{5(4\epsilon(-11+u)+u-u^{2})}{99+110\epsilon(1-u)+u(32-31u)}
\end{eqnarray}
The positive expected value is simply and intuitively due to changing
the weights $p_{ij}$ in the weighted sum in Eq. (\ref{eq:3}), which
shows the tacit dependence between the simple games. 

\begin{figure}[p]
\begin{center}
\epsfig{file=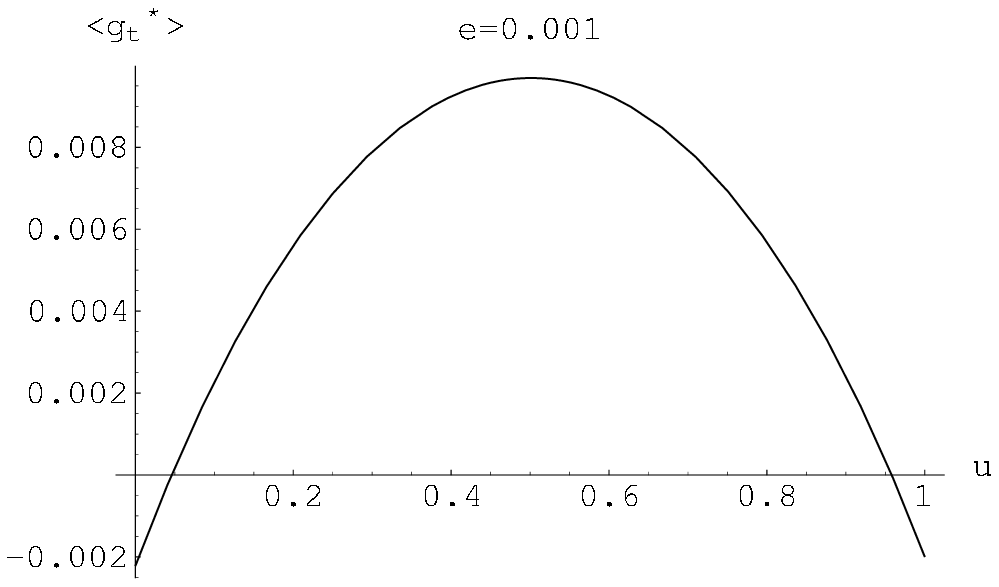,width=8cm}
\end{center}
\caption{The expected value of the mixed Parrondo game, for different values of the mixing parameter $u$. The expected value can be negative as well as positive, and that when $\epsilon=0.001$ the optimal game value is obtained for $u = 0.5012$.} 

\begin{center}
\epsfig{file=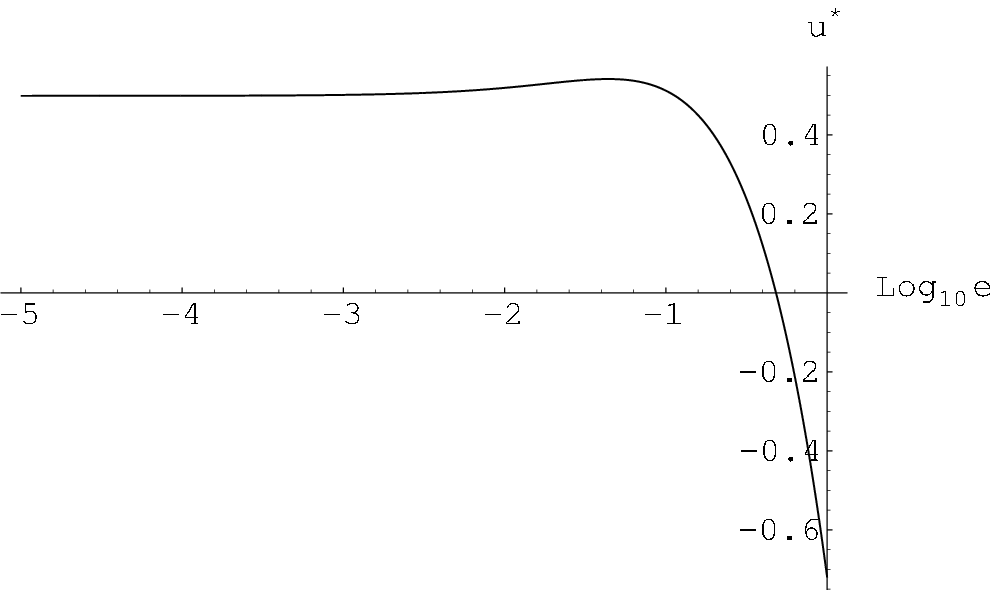,width=8cm}
\end{center}

\caption{The optimal mixing parameter $u^{*}$. Since $u^*$ decreases
rapidly for large values of $\epsilon$, the history-dependent game
outperforms the biased coin-flipping game in that region. For small
values, the optimal mixing value is $0.4987$.  For
$\log_{10}\epsilon>-0.3$ the plot shows that the game performance is
maximized by $u<0$.  In this region the optimal mixing strategy is
$u=0$, since probabilities are positive reals.}


\begin{center}
\epsfig{file=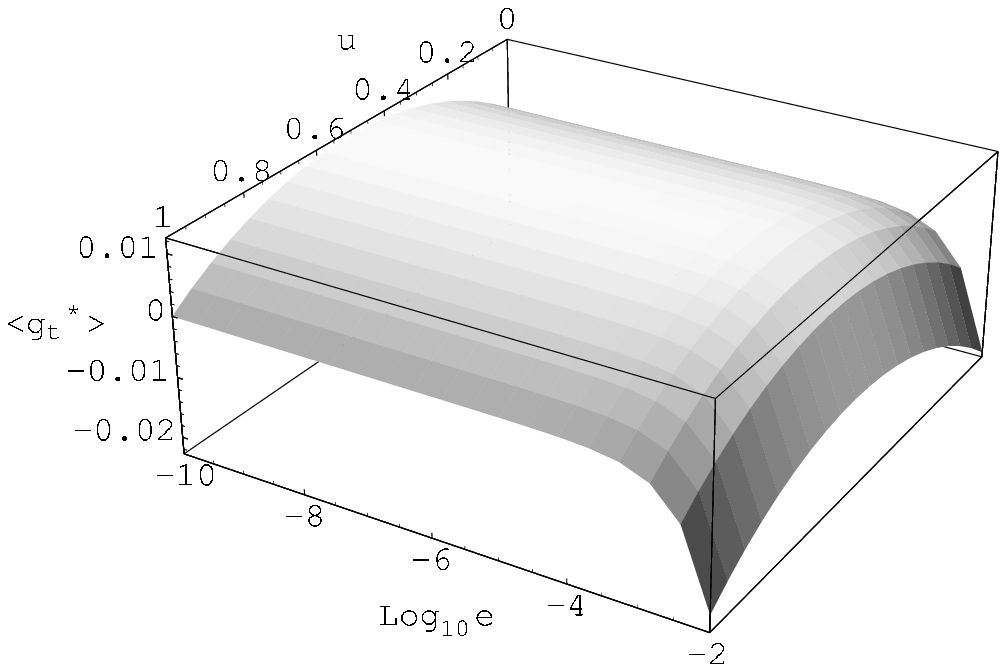,width=8cm}
\end{center}
\caption{The expected value of the mixed Parrondo game, as a function of $u$ and $\epsilon$.} 
\end{figure}

\section{\label{sect:opt}Optimal Mixing Strategies}

Harmer and Abbott~\cite{haab02} have experimentally studied a 
parameter for the probability of playing the simple games in
a Parrondo game, in order to maximize the capital of the player.
The optimal state-independent strategy $u^{*}$ is found by maximizing
$\left\langle g_{t}^{*}\right\rangle $ over $u\in [0,1]$. For $\epsilon =1/1000$,
$u^{*} \approx 0.5012$ (see Figures 1 through 3). In a similar manner, the optimal state-dependent
strategies can be calculated by defining $\left\langle g_{t}^{*}\right\rangle $
as a function of the conditional probabilities $p_{i|jk\cdots }^{*}$.

\section{\label{sect:orig}Parrondo's Original Game has Positive Expected Value}

In Parrondo's original game (see~\cite{haab99}), the positive game
outcome again depends on the tacit game interaction with a state
parameter, in this case the accumulated capital.  It mixes two simple
losing games with equal probability, hence $u=1/2$ in our notation.
The game outcome at time $t$ is $g_{t}\in \{-1,1\}$.  The winning
probabilities in one of the games depend on the accumulated capital
$C_{t}=C_{t-1}+g_{t-1}$.  The conditional transition probabilities are
given by
\begin{eqnarray} 
p_{1|;0} & = & \frac{1}{2}P+\frac{1}{2}P_{1}\qquad
p_{1|;1}=p_{1|;2}=\frac{1}{2}P+\frac{1}{2}P_{2}\label{eq:gametree}
\end{eqnarray}

\begin{equation}
P=1/2-\epsilon ,\qquad P_{1}=1/10-\epsilon ,\qquad P_{2}=3/4-\epsilon \label{eq:values}\end{equation}
where we use the notation\[
p_{ijk\cdots ;l}^{t}=\textrm{Prob}[g_{t}=i,g_{t-1}=j,g_{t-2}=k,\cdots ,C_{t}\equiv l\ ({\rm mod}\ M)]\]
 and skip the $t$ for the ergodic transition probabilites, and denote
conditional probability $p_{i\cdots |j\cdots }=p_{i\cdots }/p_{j\cdots }$.
Hence, $p_{1|;0}$ is the probability of winning when the capital
$C\equiv 0\, ({\rm mod}\  M)$. For $M=3$ we observe that

\begin{equation}
p_{;0}=p_{-1;1}+p_{1;2}\qquad p_{;1}=p_{-1;2}+p_{1;0}\qquad p_{;2}=p_{-1;0}+p_{1;1}\label{eq:modprob}\end{equation}
and since\begin{equation}
p_{i;j}=p_{i|;j}p_{;j}\label{eq:bayes}\end{equation}
we can solve for the unknown $p_{;i}$ in the linear system (\ref{eq:gametree}),
(\ref{eq:values}), (\ref{eq:modprob}), and (\ref{eq:bayes}) which
for $\epsilon =1/1000$ has the solution\begin{equation}
p_{;0}=\frac{95672}{276941}\qquad p_{;1}=\frac{10046}{39563}\qquad p_{;2}=\frac{110947}{276941}\label{eq:sol}\end{equation}
The unconditional probability of winning is\begin{eqnarray}
p_{i;} & = & \sum _{0\leq j<M}p_{i|;j}p_{;j}\label{eq:winprob}
\end{eqnarray}
and hence, from (\ref{eq:gametree}), (\ref{eq:values}), (\ref{eq:sol}),
and (\ref{eq:winprob}), the probability of winning is \[
p_{1;}=\frac{17714723}{34617625} \approx 0.5117\]
and therefore
\begin{eqnarray*}
\left\langle g_{t}\right\rangle  & = & p_{1;}-(1-p_{1;}) \approx 0.0234
\end{eqnarray*}

\section{\label{sect:outro}Conclusion}
Our analysis of Parrondo games sheds light on the phenomenon known as the Parrondo effect, or Parrondo's paradox. We have shown in a simple way the interplay between the Parrondo game constituents. 
While the expected values of the games had already been analyzed, the crucial observation
is that all Parrondo games are state-dependent, and thus when mixing Parrondo games one game
may alter the game state of another game. Finally, we have provided the optimal value for the mixing parameter in mixed Parrondo games.

\section*{Acknowledgements}
Rasmusson was funded through his position at the Intelligent Systems Lab at SICS, and Boman was funded through the Vinnova project Accessible Autonomous Software. The authors wish to
thank Derek Abbott and the anonymous reviewers for commenting on an earlier draft.


\end{document}